# Non-physical consequences of the Muffin-tin – type intra-molecular potential

M. Ya. Amusia[1, 2] and A. S. Baltenkov[3]

[1]Racah Institute of Physics, the Hebrew University, Jerusalem 91904, Israel
[2]Ioffe Physical-Technical Institute, St.-Petersburg 194021, Russia
[3]Arifov Institute of Electronics, Tashkent, 100125, Uzbekistan

**Abstract**

We demonstrate using a simple model that in the frame of muffin-tin – like potential non-physical peculiarities appear in molecular photoionization cross-sections that are a consequence of "jumps" in the potential and its first derivative at some radius. The magnitude of non-physical effects is of the same order as the physical oscillations in the cross-section of a two-atomic molecule. The role of the size of these "jumps" is illustrated by choosing three values of it.

The results obtained are connected to the studied previously effect of non-analytical behavior as a function of $r$ of the potential $V(r)$ acting upon a particle on its photoionization cross-section. In reality, such potential has to be analytic in magnitude and first derivative function in $r$. Introduction of non-analytic features in model $V(r)$ leads to non-physical features in the corresponding cross-section – oscillations, additional maxima etc.

## I. Introduction

Almost all nuclei, atoms, molecules, clusters and macroscopic bodies consist of a number of particles. Since the respective Dirac and Schrödinger equations that take into account the inter-particle interactions describing these objects cannot be solved precisely enough by using even the most advanced computers, approximations are necessary and simplifications are inevitable.

One of the most popular simplifying approaches is the so-called *mean field* approximation. In its frame it is assumed that interparticle interaction can be taken into account sufficiently accurate by a choice of mean field acting upon constituent particles – nucleon in nuclei, electrons in atoms and molecules etc.

Best known is the Hartree-Fock (HF) approximation that is good and reasonable easy solvable for atoms and nuclei. But even for atoms that are not too complex many-particle objects such a simplified version of HF as Hermann – Skillman potential [1] is used. For molecules very often muffin-tin type potential [2] is employed. In solid bodies, along with muffin – tin potentials (MTP)[3] very often corrections are introduced to improve the mean potentials that eliminate self-action of electrons (see, e.g. [4]). In all mentioned approaches the one-particle potential $V(r)$ as a function of the distance from the center of the considered system are non-analytic: their magnitudes and/or first derivatives have discontinuities.

As it was demonstrated long ago in connection to the Hermann-Skillman potential [5][1], the discontinuities manifests themselves in spurious, i.e. entirely non-physical oscillations in e.g. photoionization cross-section. This effect was reanalyzed later in a number of papers [6-8] leading to essentially the same results. Quite recently it was demonstrated that crude elimination of the self-action effect dramatically the photoionization cross-section of clusters [9].

In this paper we will concentrate on the muffin-tin potential as it is applied to photoionization processes of two-atomic molecules.

---

[1] It was also demonstrated in [5] that the finite range of numerical integration over radiuses that substitutes infinite limits serve also as a discontinuity that is indeed reflected by non-physical oscillations in the photoionization cross-section at high photon energies.

## 2. Muffin-tin potential

As is said in the Introduction, muffin-tin potential (MTP) is the theoretical construction that is widely used in the solid state and molecular physics. Within the framework of this model, the potential of the multi-atomic systems is represented as a cluster of non-overlapping spherical potentials centered on the atomic sites. In the space between the atomic spheres the potential assumed to be a constant. In solid-state physics MTP covers all space. Therefore, there is no question on the potential behavior beyond the microscopic body [10]. For the molecular case the situation is quite different. MTP here is created by a finite number of the atomic spheres, and it is impossible to neglect the existence of the molecular boundary. The adaptation of MTP for this case consists in introducing a molecular sphere [2], which surrounds all atoms that form the molecule, thus creating a sort of a resonator for the outgoing photoelectron waves.

Outside MTP sphere the photoelectron experience pure Coulomb potential of the vacancy created in photoionization process. The center of this Coulomb potential is the center of the molecular sphere. This adaptation of MTP (MTP-model) was suggested in [2] and it is currently widely used in calculations of molecular continuum wave functions and molecular photoionization (see e.g. [11 - 15] and references therein).

It is easy to see, that in MTP-model the potential of the charged molecule on surfaces of molecular sphere "jumps", forming a potential barrier. Partial reflection of the photoelectron wave emitted from the center of an atom by this potential barrier inevitably leads to alteration of the wave function inside atomic sphere where the matrix element of photo-effect is formed. In spherically - symmetric cases [16 -18] this phenomenon leads to oscillations in frequency dependences of photoionization cross-sections of atoms located inside the potential cavity sphere. The amplitude and period of these oscillations strongly depends on radius of the "resonator" and on coefficient of reflection of an electronic wave by its walls.

The physical origin of these resonances is the same as in the extended x-ray absorption fine structure (EXAFS) [19, 20]. The Kronig's short-range order theory [21] explains EXAFS by the modulations of the photoelectron wave function in the final state. The difference in the two cases is that the source of the reflected waves in the EXAFS phenomenon is the nearby atoms of the crystal, while in the MTP-model the source of backscattering waves is potential barrier that also modifies the frequency dependencies of molecular photoionization.

Hence, in MTP-model [2, 11] natural diffraction effects caused by multi-center character of the molecular photoionization problem somehow combine with artificial barrier effects. The MTP sphere is an imagined design, and its radius $R_m$ is chosen, generally speaking, arbitrarily: any sphere with radius $R > R_m$ also can be considered as MTP border. Thus, it is clear, that the results received in MTP frame, always include some uncertainty, connected to the choice of molecular sphere radius. This radius determines also the height of a potential barrier on the border of MTP sphere. As far as we know, these features of MTP-model never attract attention, and artificial barrier effects were not investigated at all.

It is the purpose of the present article to clarify to some extend the role of these features in formation of continuum molecular wave functions. Using the example of spherical models we analyze, what are the consequences in calculating continuum wave functions and photoionization cross-sections of "jumps" in potential and it's derivative.

The plan of the article is as follows. In Section 3 we analyze types of peculiarities in MTP potential. Then in Section 4 we consider spherically-symmetric atom-like system with a potential containing similar feature, and derive the equations for calculating continuum wave functions in this model problem. The obtained general formulas will be used in Sections 5 and 6 for numerical calculations of the wave functions. The parameter typical for MTP-model will be used in these calculations.

### 3. Details of the MTP model

According to paper [2] "The molecular field is defined by a potential consisting of three types of regions, defined by a set of non-overlapping spheres". There are the spherical regions $I_1$ and $I_2$ where atomic potentials are assumed to be spherically symmetric (Fig.1.) Outside these spheres but inside the molecular one (region II) the molecular potential assumed to be a constant. Outside the molecular sphere, for ionized molecules the molecular potential is the Coulomb potential originated from the molecular sphere center.

Consider the behavior of the potential in which moves the photoelectron, eliminated after photon absorption from a deep level of, for example, first atom $I_1$. If to move a trial charge along the line 1, as shown in Fig.1, counting coordinate $r$ from a nucleus of this atom we have the following qualitative picture. Inside atomic sphere $I_1$ the potential is close to potential of a positive ion. On the border of atomic sphere the curve $V(r)$ experience a break at transition into a constant. Between points A and B on the line 1 the molecular potential is constant. Then (at the point B) MTP potential "jumps" at the molecular sphere, changing from $V(R_a)$ to $-1/R_m$ [2]. Examining other lines of a trial charge motion, we conclude, that the MTP-model potential looks like a step of constant height, but various width. So for a trajectory 2 that corresponds to nearest to atomic center point of connection of atomic and molecular spheres, the width of this potential step is equal to zero.

Thus, in MTP-model we have two following features in potential: a jump of the potential's derivative on surfaces of atomic spheres and jump of the potential magnitude on the surface of the molecular sphere. The role of the potential jogs was recently analyzed also in [22] in connection with photoionization of so-called hollow atoms (HA). It was shown that the amplitude of oscillations in the energy dependence of the photoionization cross section of HA is extremely sensitive to a magnitude of the potential derivative discontinuity. As far as the effect of potential jumps on the particle wave function is concerned, it is well studied in connection to the square well potential (see e.g. in [23, 24]).

It is easy to estimate the size of potential jump if to consider atomic ion $V(r)$ potential as pure Coulomb. In this case the height of the potential barrier is $\Delta V = 1/R_a - 1/R_m$. For radiuses of spheres $R_a \approx 1$ also $R_m \approx 2$ it means that $\Delta V \sim$ Ry. Thus, the height of the spherical potential barrier surrounding a molecule in MTP-model is not small on atomic scales[3].

In photoionization of a deep atomic level the electronic wave, being distributed from the center of sphere $I_1$, passes above this barrier. Its reflection leads to change of the amplitude of the wave function near the atomic nucleus by $F(k)$, and, hence, to change in $|F(k)|^2$ of photoionization cross-section (here $k$ is the photoelectron's linear momentum). Due to connection between electron wave-function oscillations the potential resonator and outside of it, the amplitude $F(k)$ as a function of momentum $k$, looks like a fading sinusoid [25]. At $k^2/2 \gg \Delta V$ the potential jump $\Delta V$ is insignificant and therefore $F(k \to \infty) \to 1$. Oscillating behavior $F(k)$ is obvious already from general arguments, but the magnitude of $F(k)$ can be established only by numerical calculation. Note, that the oscillations $F(k)$ in considered spherically - symmetric systems with similar potential barriers are able to change the photoionization cross-section of an atomic deep level near threshold by several times.

The diffraction effects in molecular photoionization are approximately described by the function $F_{dif}(k) = (1 + \sin kR / kR)$ that multiplies photoionization cross-section of isolated atom ionization [26], where *R* is the inter-atomic distance in a two-atomic molecule. At the threshold ($k = 0$) this function increases the photoionization cross-section by a factor 2. Hence, the "hand-made" barrier effects in MTP-model are comparable or even surpass the real diffraction ones, and therefore the research of those MTP effects with the aim of their elimination is essential.

---

[2] Atomic system of units is used in this paper

[3] Note that the potential well depth for $C_{60}$ is two times smaller, $\Delta V \approx 8 eV$ [20, 21]. However, the presence of this potential resonator alters radically the frequency dependence of photoionization cross-section of an atom, located inside $C_{60}$ in A@$C_{60}$.

## 4. Spherically-symmetric potential barrier

Consider the model problem, allowing analyzing the influence of a potential step on continuous spectrum wave functions. Let us calculate the wave functions of a hydrogen-like atom with the nuclear charge Z, surrounded by a potential barrier. It is obvious, that because of spherical symmetry of considered atom-like system this calculation is reduced to obtaining the radial part of electron wave function $P_{kl}(r)$ with linear momentum *k* in continuous spectrum and the angular moment *l*. The potential in which the electron moves is defined as follows:

$$
\begin{aligned}
V(r) &= -Z/r - \Delta V & (r \leq R_a) \\
V(r) &= -Z/R_a - \Delta V & (R_a \leq r \leq R_m) \\
V(r) &= -Z/r & (r > R_m)
\end{aligned}
\quad (1)
$$

Here $\Delta V$ is the height of potential jump at the point where atomic $I_1$ and molecular spheres touch each other. From potential (1) it is obvious, that wave function $P_{kl}(r)$ is a combination of Coulomb functions and the spherical Bessel functions "sewed" at points $R_a$ and $R_m$.

Let $u_{kl}(r)$ and $v_{kl}(r)$ denote regular and irregular at $r = 0$ Coulomb wave functions with asymptotic behavior:

$$
\begin{aligned}
u_{kl}(r) &\approx \sin(kr + \frac{Z}{k}\ln 2kr - \frac{\pi l}{2} + \delta_l) \\
v_{kl}(r) &\approx -\cos(kr + \frac{Z}{k}\ln 2kr - \frac{\pi l}{2} + \delta_l)
\end{aligned}
\quad (2)
$$

where $\delta_l(k) = \arg \Gamma(l + 1 - i\eta)$ is the Coulomb phase shift. Then in the first region of *r* the radial part of wave function is:

$$P_{\kappa l}(r) = F_l(k) u_{\kappa l}(r) \quad (r \leq R_a). \quad (3)$$

The wave function in this area differs from the regular solution of the Schrödinger equation with potential (1) $u_{kl}(r)$ only by an amplitude multiplier $F_l(k)$. The electron wave vector $\kappa$ in this area is defined by the equation $\kappa^2/2 = k^2/2 + \Delta V$.

In the second area of *r*

$$P_{kl}(r) = A j_l(qr) - B n_l(qr) \quad (R_a \leq r \leq R_m) \quad (4)$$

Here $j_l(qr)$ and $n_l(qr)$ are the spherical Bessel functions [27] multiplied by *qr*, with asymptotic behavior

$$j_l(qr) \approx \sin(qr - \frac{\pi l}{2}), \quad n_l(qr) \approx -\cos(qr - \frac{\pi l}{2}), \quad (5)$$

The electron wave vector $q$ in functions (4) and (5) is defined by the equation $q^2/2 = k^2/2 + Z/R_a + \Delta V$.

In the third area of distances

$$P_{kl}(r) = u_{kl}(r)\cos\Delta_l - v_{kl}(r)\sin\Delta_l \quad (r > R_m), \tag{6}$$

Here $\Delta_l$ is additional phase shift, which is acquired by the electron wave function while passing the potential step.

At points $r = R_a$ and $r = R_m$ the functions and their derivatives should be sewed. It leads to the following system of four equations

$$\begin{aligned}
F_l u_{kl}(R_a) &= A j_l(qR_a) - B n_l(qR_a), \\
F_l u'_{kl}(R_a) &= A j'_l(qR_a) - B n'_l(qR_a), \\
A j_l(qR_m) - B n_l(qR_m) &= u_{kl}(R_m)\cos\Delta_l - v_{kl}(R_m)\sin\Delta_l, \\
A j'_l(qR_m) - B n'_l(qR_m) &= u'_{kl}(R_m)\cos\Delta_l - v'_{kl}(R_m)\sin\Delta_l.
\end{aligned} \tag{7}$$

The mark prime (') means differentiation on $r$. Here unknown are coefficients $A$, $B$, $F_l$ and $\Delta_l$. All of them are functions $k$ and $l$.

To find $F_l$ and $\Delta_l$ we multiply the first equation by $n'_l(qR_a)$, and the second by $n_l(qR_a)$ then deducting the first equation from the second. The third equation is multiplied by $u'_{kl}(R_m)$, and the fourth by $u_{kl}(R_m)$ and them from each other. Finally we receive

$$\tan\Delta_l(k) = \frac{W_1 W_4 - W_2 W_3}{W_1 W_6 - W_2 W_5} \tag{8}$$

Here

$$\begin{aligned}
W_1 &= u_{kl}(R_a) j'_l(qR_a) - u'_{kl}(R_a) j_l(qR_a) \\
W_2 &= u_{kl}(R_a) n'_l(qR_a) - u'_{kl}(R_a) n_l(qR_a) \\
W_3 &= u_{kl}(R_m) j'_l(qR_m) - u'_{kl}(R_m) j_l(qR_m) \\
W_4 &= u_{kl}(R_m) n'_l(qR_m) - u'_{kl}(R_m) n_l(qR_m) \\
W_5 &= v_{kl}(R_m) j'_l(qR_m) - v'_{kl}(R_m) j_l(qR_m) \\
W_6 &= v_{kl}(R_m) n'_l(qR_m) - v'_{kl}(R_m) n_l(qR_m)
\end{aligned} \tag{9}$$

For the amplitude $F_l(k)$ the following two equivalent equations are received

$$\begin{aligned}
F_l(k) &= \frac{qk\sin\Delta_l(k)}{W_1 W_4 - W_2 W_3}, \\
F_l(k) &= \frac{qk\cos\Delta_l(k)}{W_1 W_6 - W_2 W_5}.
\end{aligned} \tag{10}$$

Let us note that the amplitude $F_l(k)$ is always positive, since only in this case the continuum wave function near zero in Eq. (3) behaves as it should be in a centrally symmetric field, i.e. as $P_{\kappa l}(r \to 0) \sim r^{l+1}$.

The case of zero-width step ($R_a = R_m$) requires special consideration. In this case $F_l$ and $\Delta_l$ are determined from equality of logarithmic derivative of Coulomb function (3) and (6) at this point. Putting in (7) $R_a = R_m$ and equating the left parts of first two equations to the right parts of the second pair of these equations, and repeating calculations similar to the previous, one obtains for the phase $\Delta_l$ and amplitude $F_l$ the following expressions

$$\tan \Delta_l(k) = \frac{W_7}{W_8}, \tag{11}$$

$$F_l(k) = \frac{\cos \Delta_l(k)}{u_{\kappa l}(R_a)}[u_{kl}(R_a) - v_{kl}(R_a)\tan \Delta_l(k)]. \tag{12}$$

Here functions $W_7$ and $W_8$ are given by expressions

$$\begin{aligned} W_7 &= u_{\kappa l}(R_a)u'_{kl}(R_a) - u'_{\kappa l}(R_a)u_{kl}(R_a), \\ W_8 &= u_{\kappa l}(R_a)v'_{kl}(R_a) - u'_{\kappa l}(R_a)v_{kl}(R_a). \end{aligned} \tag{13}$$

We apply the obtained formulas in numerical calculations of amplitudes $F_l(k)$ for various heights and width of the potential step.

## 5. Dependence of the amplitude $F_l(k)$ on the height of barrier

Let us consider again the case of zero width potential step that is the case $R_a = R_m = 1$. Results of calculation of the amplitude $F_0(k)$ with orbital momentum $l = 0$ using formulas (12) and (13) at various heights of the barrier $\Delta V = 0.01$, 0.1 and 0.5 a.u. are presented in Fig. 2. On insert in this figure the potential of the considered system $V(r)$ is depicted, formed by two Coulomb potentials

$$\begin{aligned} V(r) &= -Z/r - \Delta V \quad (r \leq R_a), \\ V(r) &= -Z/r \quad (r > R_a). \end{aligned} \tag{14}$$

In calculations presented below we put $Z = 1$. We consider the size of $\Delta V$ and the width of the barrier as parameters.

Functions $F_0(k)$ in Fig. 2 look like periodically changing functions with almost identical period $\Delta k \approx 3.6$. Deviations from perfect periodic behavior are observed only at small electron kinetic energies. The amplitudes of oscillations decrease with growth of electron momentum and curves reach its asymptotic value $F_0 = 1$. With increase of barrier height $\Delta V$ the amplitude of oscillations, as one would expect, grows.

In Fig. 3 the amplitude factor $F_1(k)$ for wave function with the orbital moment $l = 1$ is represented. Here $V(r)$ along with (14) includes also the centrifugal potential $l(l+1)/2r^2 = 1/r^2$ for $l = 1$. This potential supersedes wave function from area where there is a jump of potential $V$. Therefore at small $\Delta V = 0.01$ the presence of a barrier is insignificant and $F_1(k) \approx 1$ in all the considered interval of electron momentum. As in Fig. 2, the amplitude's factor oscillates with practically the same period $\Delta k \approx 3.3$. The difference is that all curves on Fig. 2 change similarly whereas the curves corresponding to $\Delta V = 0.1$ and 0.5 at $k \approx 7.4$ are strictly in anti-phase.

## 6. Dependence $F_l(k)$ on parameters of a potential step

The potential (14) corresponds in MTP-model to a case when an electron moves in a continuous spectrum along a trajectory 1 for which the width of a potential step is equal to zero. It is obvious, that in a molecule described by MTP-model, a spherical electronic wave, being distributed in all directions from the center of atomic sphere $I_1$, experience reflection from a potential step of variable width. It corresponds to atom-like system with potential (1). For numerical calculations we shall choose the jump of potential on the border of molecular sphere $\Delta V$ so that the average height of the potential barrier on all possible trajectories of electron movement was equal to $\Delta \overline{V} = 0.5$. We shall define it as an arithmetic average of the minimal and maximal heights of a potential step. For a fixed atomic radius $R_a = 1$, the molecular sphere radius is three times more, $R_m = 3$. Therefore, to fulfill the condition $\Delta \overline{V} = 0.5$ it is necessary to put in (1) potential $\Delta V = 1/6$.

The results of $F_l(k)$ calculation at two different electron orbital moments are resulted in Fig. 4 and 5. On inserts in these figures the potential $V(r)$ calculated according (1) is presented. The following width of potential step $w = R_m - R_a = 0.5, 1.0, 1.5$ and $2.0$ were considered. Comparison with curves in Fig. 2 and 4 shows, that the general behavior of amplitude $F_0(k)$ is similar. In both cases they are fading oscillating functions. But oscillating curves in Fig. 4, as curves in Fig. 2, have different periods. This is due to change of the potential radius $R_m$. Besides final width of a potential step, to be exact, presence of both types of features of potential, namely: jumps of the derivative $dV/dr$ and of the potential $V(r)$, modifies the form of curves, making them smoother.

## 7. Conclusions

As it was mentioned above, the photoionization cross-section is proportional to the square of $F_l(k)$. Therefore, non-physical barrier effects in the considered model (14) change the photoionization cross-section at resonant electron energies more than by an order of magnitude that is considerably stronger than variations due to the diffraction factor $F_{dif}(k) = (1 + \sin 2kR_a / 2kR_a)$, corresponding to a molecule, formed by two identical atoms (Fig. 1). For such molecule the inter-atomic distance is $2R_a$. The diffraction period is close to $\Delta k \approx 3.14$ i.e. to the periods of functions $F_0(k)$ and $F_1(k)$. It means that barrier and diffraction effects have similar periodic structure. But the maximal squares of barrier functions essentially surpass $F_{dif}(k)$.

## Acknowledgements

This work was supported by the Hebrew University intramural fund and Uzbek Foundation Award ФА-Ф4-Ф100.

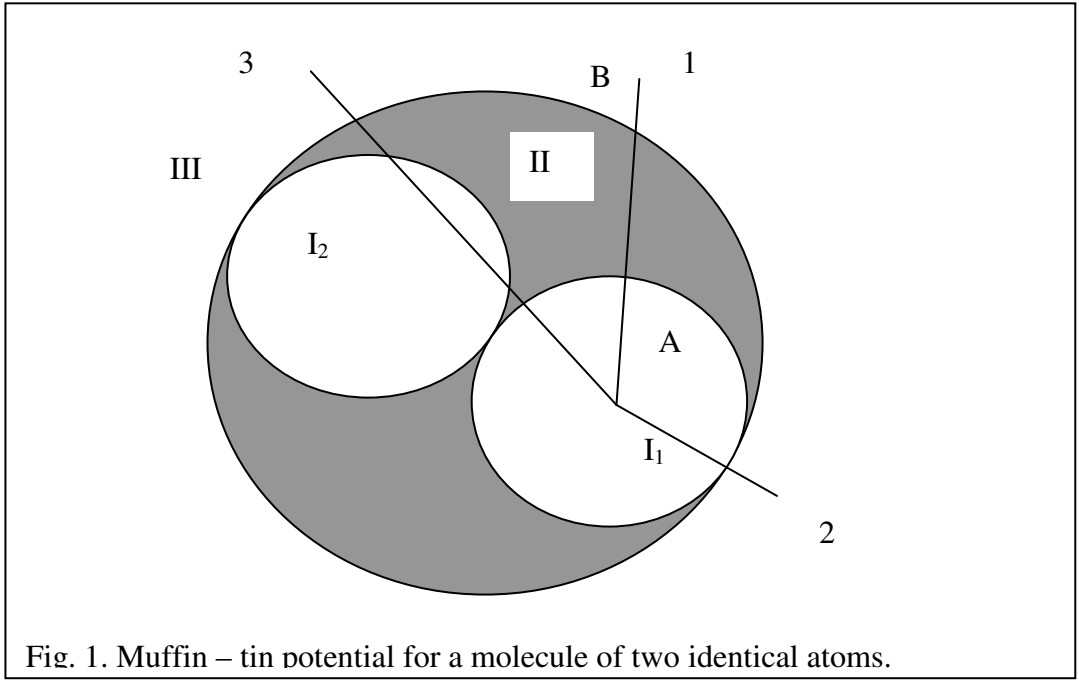

Fig. 1. Muffin – tin potential for a molecule of two identical atoms.

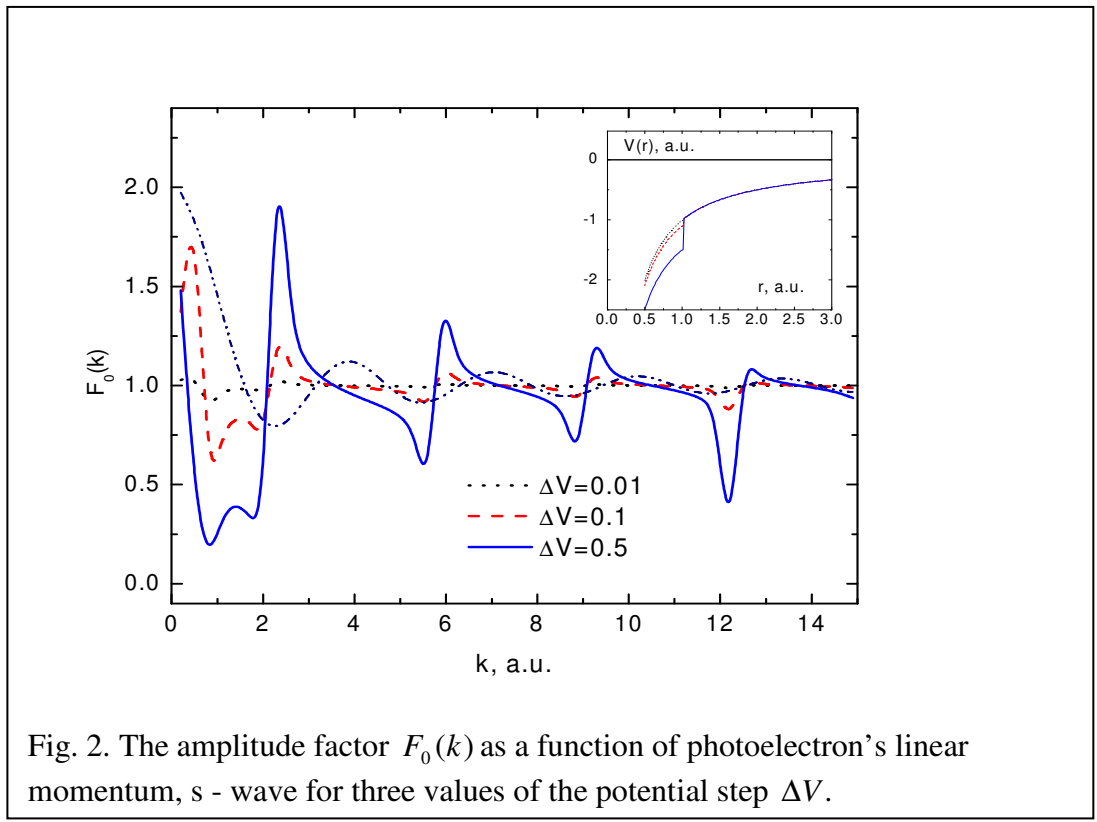

Fig. 2. The amplitude factor $F_0(k)$ as a function of photoelectron's linear momentum, s - wave for three values of the potential step $\Delta V$.

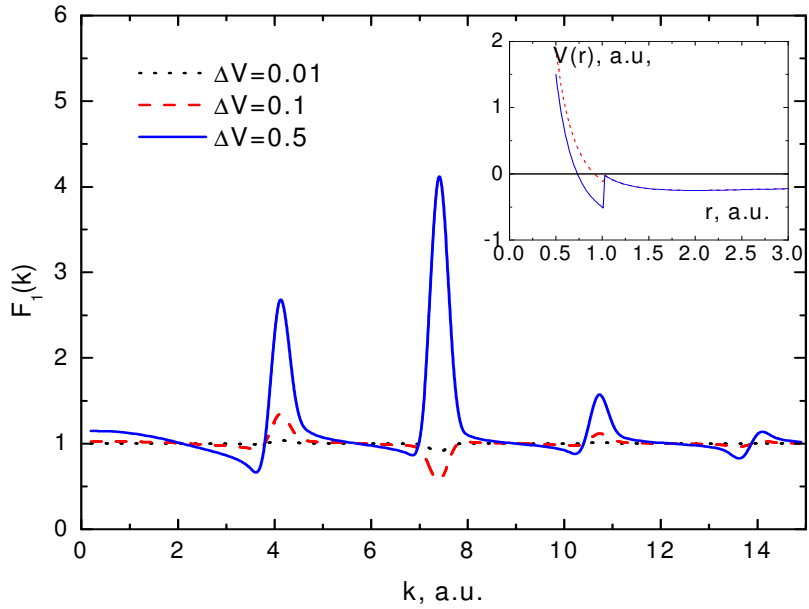

Fig. 3. The amplitude factor $F_1(k)$ as a function of photoelectron's linear momentum, s-wave for three values of the potential step $\Delta V$.

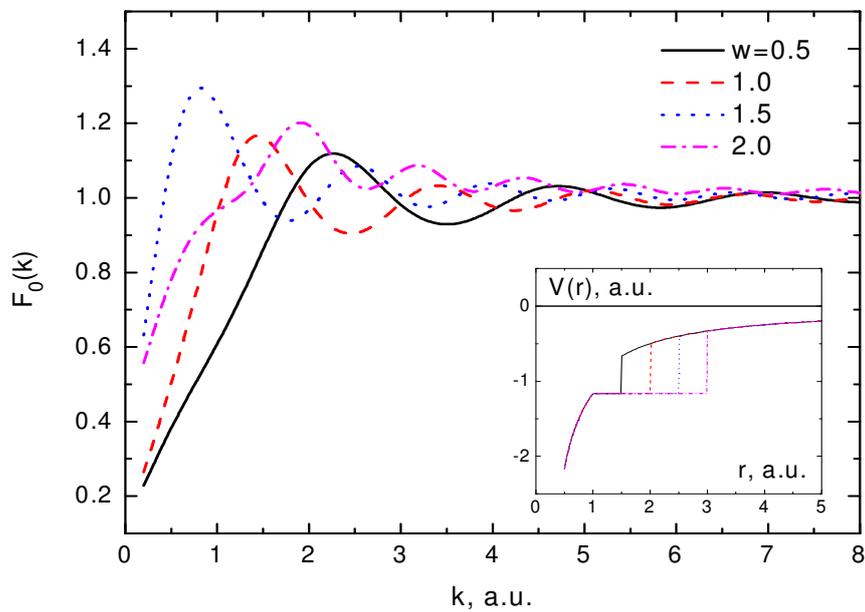

Fig. 4. The amplitude factor $F_0(k)$ as a function of photoelectron's linear momentum, s-wave, for four values of the potential step $\Delta V$ width w.

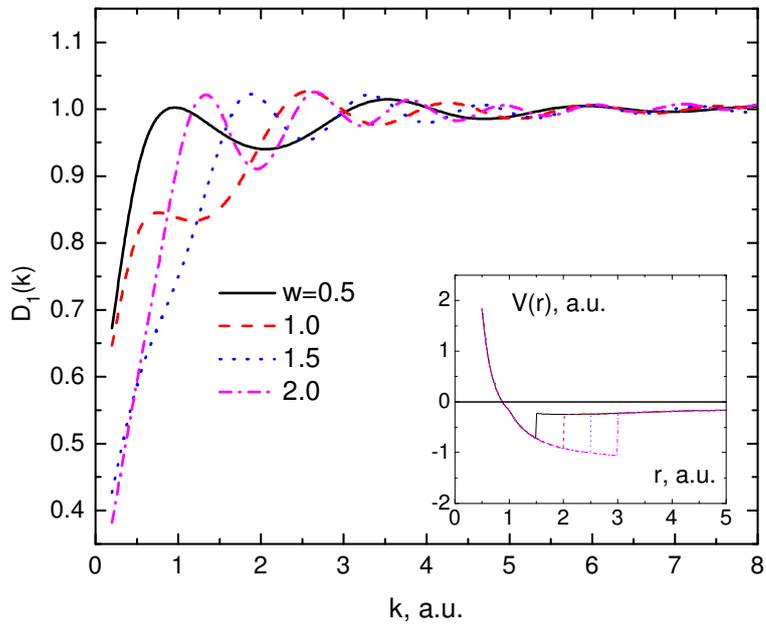

Fig. 5. The amplitude factor $F_1(k)$ as a function of photoelectron's linear momentum, p-wave, for four values of the potential step $\Delta V$ width w.